\documentclass[conference]{IEEEtran}
\IEEEoverridecommandlockouts
\usepackage{cite}
\usepackage{amsmath,amssymb,amsfonts}
\usepackage{algorithmic}
\usepackage{graphicx}
\usepackage{textcomp}
\usepackage{xcolor}
\usepackage{amsmath,amsfonts}
\usepackage{algorithmic}
\usepackage{algorithm}
\usepackage{array}
\usepackage{textcomp}
\usepackage{stfloats}
\usepackage{url}
\usepackage{verbatim}
\usepackage{graphicx}
\usepackage{cite}
\usepackage{hyperref}
\def\BibTeX{{\rm B\kern-.05em{\sc i\kern-.025em b}\kern-.08em
    T\kern-.1667em\lower.7ex\hbox{E}\kern-.125emX}}
\begin{document}

\title{Semantic Communications for 3D Human Face  Transmission with Neural Radiance Fields\\
\begin{center} 
    \Large(\emph{Invited Paper})
\end{center}
}

\author{Guanlin Wu$^1$, Zhonghao Lyu$^1$, Juyong Zhang$^2$, and Jie Xu$^{1}$\\
$^1$School of Science and Engineering (SSE) and Future Network of Intelligence Institute (FNii),\\ The Chinese University of Hong Kong (Shenzhen), Shenzhen, China\\
$^2$School of Mathematical Science, University of Science and Technology of China, Hefei, China\\
E-mail: guanlinwu1@link.cuhk.edu.cn,~zhonghaolyu@link.cuhk.edu.cn,\\juyong@ustc.edu.cn,~xujie@cuhk.edu.cn

\thanks{The work was supported in part by the National Natural Science Foundation of China under grants No. U2001208, the Basic Research Project No. HZQB-KCZYZ-2021067 of Hetao Shenzhen-HK S\&T Cooperation Zone, the National Natural Science Foundation of China under grants No. 92267202, No.62122071, and No.62272433, the Shenzhen Fundamental Research Program under grant No. JCYJ20210324133405015, the Guangdong Provincial Key Laboratory of Future Networks of Intelligence under grant No. 2022B1212010001, the National Key R\&D Program of China with grant No. 2018YFB1800800, the Shenzhen Key Laboratory of Big Data and Artificial Intelligence No. ZDSYS201707251409055, the Key Area R\&D Program of Guangdong Province with grant No. 2018B030338001, and the Guangdong Major Project of Basic and Applied Basic Research (No. 2023B0303000001). J. Xu is the corresponding author.} 

}

\maketitle

\begin{abstract}
This paper investigates the transmission of three-dimensional (3D) human face content for immersive communication over a rate-constrained transmitter-receiver link. We propose a new framework named NeRF-SeCom, which leverages neural radiance fields (NeRF) and semantic communications to improve the quality of 3D visualizations while minimizing the communication overhead. In the NeRF-SeCom framework, we first train a NeRF face model based on the NeRFBlendShape method, which is pre-shared between the transmitter and receiver as the \emph{semantic knowledge base} to facilitate the real-time transmission. Next, with knowledge base, the transmitter extracts and sends only the essential semantic features for the receiver to reconstruct 3D face in real time. To optimize the transmission efficiency, we classify the expression features into static and dynamic types. Over each video chunk, static features are transmitted once for all frames, whereas dynamic features are transmitted over a portion of frames to adhere to rate constraints. Additionally, we propose a feature prediction mechanism, which allows the receiver to predict the dynamic features for frames that are not transmitted. Experiments show that our proposed NeRF-SeCom framework significantly outperforms benchmark methods in delivering high-quality 3D visualizations of human faces.
\end{abstract}

\section{Introduction}
Immersive communications have been recognized as one of the key usage scenarios in forthcoming sixth-generation (6G) wireless networks, which are expected to enable abundant applications such as extended reality (XR), virtual reality (VR), teleconference, immersive gaming, and metaverse. Different from enhanced mobile broadband (eMBB) in its fifth-generation (5G) counterpart, immersive communications in 6G aim at creating a highly engaging and interactive communication environment for users \cite{Shen_2023}, which highly rely on efficient representation and transmission of three-dimensional (3D) content. For instance, the representation and transmission of 3D human faces are crucial for the realization of immersive teleconference. These tasks, however, are particularly challenging and require interdisciplinary design approaches exploiting cutting-edge communication and computation technologies.

Efficient representation of 3D content is essential for its transmission. In general, the methods for 3D representation can be categorized as explicit and implicit ones \cite{jin2023capture}. While the explicit representation methods can intuitively display 3D objects based on mesh, voxel, or point cloud, they may not capture fine details and demand substantial data storage \cite{5980567}. By contrast, implicit representation, particularly Neural Radiance Field (NeRF) \cite{nerf3503250}, has been proposed recently as an alternative method to represent 3D content by employing neural networks. In particular, NeRF utilizes only multi-view images of a scene as training data to learn neural network models, which are able to generate novel viewing images via volume rendering. Owing to the excellent capability of neural networks, NeRF demonstrates superior performance in representing the details of 3D content \cite{gao2023nerf}, and only requires lightweight neural networks with significantly reduced model sizes \cite{nerf3503250}. Therefore, it is promising to utilize NeRF for efficiently representing 3D content like human faces.

The transmission of data-intensive 3D content over bandwidth-limited communication networks is another challenge to tackle, as the 3D content generally requires significantly more data to represent as compared to the  two-dimensional (2D) content. Recently, semantic communications \cite{9955525SeCom,zhu2023pushing} have emerged as a novel communication paradigm, which allows the transmitter to extract and transmit semantic information instead of bit-level raw data, thus significantly reducing the communication overhead while preserving the quality of experience. In the literature, there have been extensive prior works investigating the semantic transmission of 2D content (see, e.g., \cite{Lyu10388062Secom, Qin9763856textSecom,ren2023knowledge}), but only some initial works studying the semantic transmission of 3D content such as point clouds \cite{wang2024taskoriented} and $360^{\circ}$ videos \cite{Le10105154WiserVR}. How to efficiently deliver 3D content based on advanced NeRF-based representation has not been studied yet. This thus motivates our investigation in this work.

This paper studies the transmission of a particular type of 3D content, i.e., 3D human faces, for immersive communication over a rate-constrained transmitter-receiver link. We propose a novel framework, namely NeRF-SeCom, which jointly exploits NeRF and semantic communications to realize efficient representation, transmission, and reconstruction of 3D human faces. In the NeRF-SeCom framework, we pre-train a face NeRF model via the NeRFBlendShape method, which is utilized as the semantic knowledge base and shared between the transmitter and receiver {\emph{a-priori}} to facilitate the real-time transmission. With knowledge base, the transmitter only needs to extract and transmit essential semantic features to the receiver for real-time 3D face reconstruction. To enhance the transmission efficiency, we further propose a novel feature selection design and feature prediction mechanism.
On the one hand, the feature selection design classifies the expression features into static and dynamic types. In such a way, over each video chunk, the static features only need to be transmitted once for all frames, and the dynamic features are transmitted over a portion of frames to adhere to the rate limitations. On the other hand, the feature prediction mechanism allows the receiver to predict the dynamic features for frames that are not transmitted. Finally, numerical results are provided to demonstrate the effectiveness of our proposed NeRF-SeCom design framework in achieving high-quality 3D human face visualization results and outperforming various benchmark schemes.

\section{System Model} \label{system}

In this paper, we consider the transmission of 3D human faces over a transmitter-receiver link to support immersive applications such as teleconferencing and gaming. Specifically, the transmitter uses a Red-Green-Blue (RGB) camera to capture the monocular video of human faces, and then sends the captured 3D human face information to the receiver. After receiving such information, the receiver needs to reconstruct the 3D human face. 

The transmission of 3D video is implemented via video streaming techniques. Specifically, we consider video chunk as the basic transmission unit, in which each chunk contains a number of $N$ video frames. Without loss of generality, we focus on the transmission of one particular video chunk, for which each video frame $n\in\{1,\ldots,N\}$ is denoted as as $\boldsymbol{I}_n \in \mathbb{R}^{H\times W \times 3}$, with $H$ and $W$ denoting the height and width of the 3D image, respectively.

In general, there are different techniques to represent and transmit the 3D human face content. In the following, we briefly introduce the general procedure, and leave the specific NeRF-based representation and semantic communication in the next section. Without loss of generality, we consider the transmission of one video chunk of $N$ frames $\boldsymbol{I}_1, \ldots, \boldsymbol{I}_N$. First, the transmitter extracts features of $\{\boldsymbol{I}_n\}_{n=1}^N$ and encodes the extracted features to bit-streams. Then, the transmitter sends the bit-streams to the receiver via properly designed transmission strategies such as modulation and coding schemes. After receiving the bit-streams, the receiver reconstructs the 3D video of human faces by first mapping the bit-streams to the facial feature information, and then utilizing the features to render the human faces as $\{\hat{\boldsymbol{I}}_n\}_{n=1}^N$ for the whole video chunk.

The real-time transmission of 3D human faces is subject to both the latency constraint and the rate constraint of the transmitter-receiver link. Let $\tau$ denote the maximum communication latency requirement for the transmission of one 3D video chunk to ensure the quality of experiences,\footnote{Notice that the end-to-end latency for 3D human face reconstruction also constrains the computation latency for feature extraction and rendering. In order to focus on the communication design, we consider the computation latency as a constant value of $\hat{\tau}$, such that the overall end-to-end latency cannot exceed $\hat{\tau} +{\tau}$. We will investigate the effect of feature extraction and rendering on computation latency in future work.} and $R$ denote the maximum communication rate from the transmitter to the receiver.\footnote{Notice that the communication rate of the transmitter-receiver link depends on various factors such as the routing between them, the resource allocation across multiple coexisting links, the channel conditions of the wireless links connecting them. Therefore, the maximum communication rate $R$ may fluctuate over time. For the ease of initial investigation, we assume that $R$ is constant in this paper, we leave the joint design of resources allocation and NeRF-SeCom in future work.} Therefore, to ensure the latency requirements, the total bits for delivering one 3D video chunk should not exceed $\tau R$. As such, our objective is to maximize the reconstruction quality of human faces over each 3D video chunk subject to the transmission capacity constraint of $\tau R$ bits.

\section{NeRF-SeCom Framework for 3D Human Face Transmission} \label{framework}

This section presents the new NeRF-SeCom framework for 3D human face transmission, which represents the 3D human face as NeRF models to facilitate the transmission. In particular, the transmitter and receiver utilize the NeRF models as the semantic knowledge base, based on which the transmitter only needs to extract and transmit only essential semantic features in real time. We also propose efficient semantic feature selection at the transmitter and semantic feature prediction at the receiver to reduce the communication overhead to meet the transmission requirements of $\tau R$ bits.

\subsection{Review of Face NeRF}
To start with, this subsection provides a brief review on the basics of NeRF, human face NeRF, and our adopted  NeRFBlendshap.
\begin{figure}
    \centering  \includegraphics[width=1.0\linewidth]{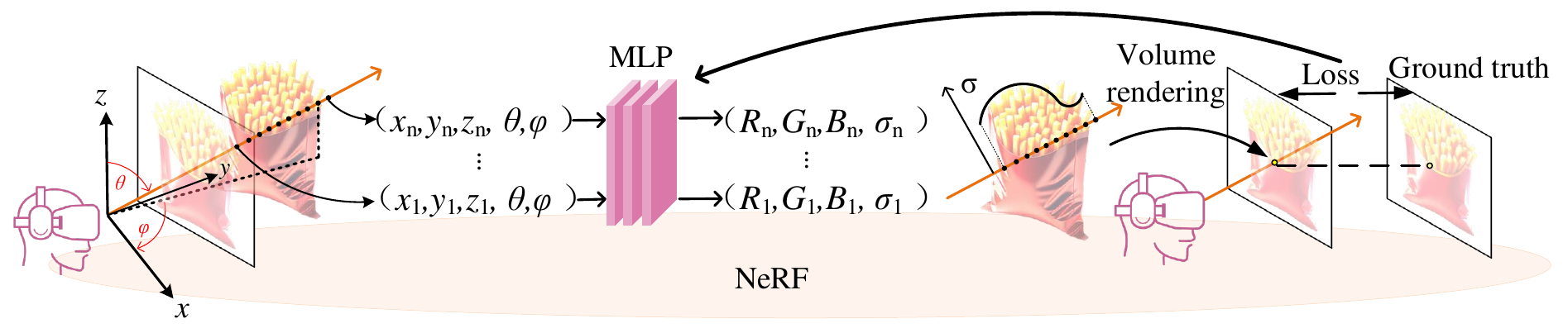}
    \caption{Basics of NeRF.}
    \label{NeRF}
\end{figure}
\subsubsection{NeRF} \label{Principle}

NeRF was originally proposed in \cite{nerf3503250}, which represents 3D content by using multi-layer perceptron (MLP) and volume rendering as shown in Fig.~\ref{NeRF}. Specifically, the MLP neural network is utilized to approximate the radiance field, which implicitly characterizes the RGB color and volume density of every volume point within the 3D scene. Accordingly, the NeRF representation is expressed as
\begin{align}
    \boldsymbol{F}_{\Theta}(\boldsymbol{x},\boldsymbol{d})\rightarrow (\sigma,\boldsymbol{c}),
\end{align}
where $\Theta $ denotes the learnable weight parameters of MLP, $\boldsymbol{x}=\left(x,y,z\right)$ represents the coordinates of volume point, $\boldsymbol{d}=(\theta,\phi)$ denotes the viewing direction unit vector, $\boldsymbol{c}=\left(R,G,B\right)$ and $\sigma$ denote the RGB color channels and the volumetric density of corresponding points, respectively. The camera ray is defined as $\boldsymbol{r}(t)=\boldsymbol{o}+t\boldsymbol{d}$, where $\boldsymbol{d}$ denotes the viewing direction and $\boldsymbol{o}$ denotes the origin. In the volume rendering phase, the colors and densities of points are accumulated as 
\begin{align} \label{volumerendering}
    I(\boldsymbol{r})=\int_{t_n}^{t_f} T(t) \sigma(\boldsymbol{r}(t)) {\boldsymbol{c}}(\boldsymbol{r}(t), \boldsymbol{d}) dt,
\end{align}
where $T(t)=\exp \left(-\int_{t_n}^t \sigma(\boldsymbol{r}(u))  d u\right)$, and $t_n$ and $t_f$ represent the nearest and the farthest volume points, respectively. Thereby, the pixels of 2D images from the given viewing direction are determined one by one. The parameters $\Theta $ of MLP are trained by using the mean square error (MSE) between sampled images and the rendered results as the loss function.

\subsubsection{Face NeRF Model}
As a particular type of 3D content, 3D human faces can be represented via combinations of a series of parameters, based on the pioneering work of 3D Morphable Model (3DMM) \cite{3DMMBlanz}. This success has led to the integration of parametric face models and NeRF, in which the facial parameters are encoded into the latent space as intrinsic attributes that can be efficiently combined with the radiance field, expressed as
\begin{align}
    \mathcal{R}_{\theta}(C,\boldsymbol{e}_{\text{face}}) \rightarrow (\sigma, \boldsymbol{c}),
\end{align}
where $\theta$ is the learnable weight parameters of MLP networks, $C$ represents the camera parameter used for rendering, and $\boldsymbol{e}_{\text{face}}$ represents the facial parameters, such as the expressions, pose, and identity. Based on this principle, various works (e.g., \cite{Wang3530753Morphable}) proposed NeRF-based face parametric models to efficiently represent facial details. However, these works require a sufficiently long time duration for rendering. 
\subsubsection{NeRFBlendShape}
To address the issue of long rendering duration, NeRFBlendShape \cite{Gao3555501nerfface} has emerged as a new 3D human face NeRF model, which adopts the multi-level hash table \cite{InstantNgp} to associate the expression coefficients for accelerating rendering. Specifically, NeRFBlendShape first utilizes the feature extraction module to extract expressions and head poses of faces. Next, these expression parameters and the sampled points of rays are associated with the hash table \cite{Gao3555501nerfface}. The querying results of the hash table and the viewing direction are used together as inputs to the MLP network. The output of MLP provides the color and density values of the sampled points. Finally, volumetric rendering is performed on the resultant color and density to obtain the facial visualization results. The NeRFBlendShape face model is expressed as 
\begin{align} \label{NeRFBlendShapemodel}
    \hat{\mathcal{R}}_\beta\left(C, \boldsymbol{e}\right)\rightarrow (\sigma, \boldsymbol{c}),
\end{align}
where $\hat{\mathcal{R}}$ represents the querying of hash table and MLP networks, $\beta$ represents the learnable weight parameters of the MLP network, and $C$ and $\boldsymbol{e}=\left\{e_1,e_2,\ldots,e_M \right\}\in \mathbb{R}^{M}$ denote the camera parameters and the expression feature coefficients, respectively. Here, $M$ denotes the total dimension of the expression features. NeRFBlendShape adopts expression as the semantic feature to represent 3D faces. \footnote{It is important to note that besides expression features, some other works incorporated additional characteristics such as identity and appearance (see, e.g., \cite{Hong_2022_CVPR}), which is left for future work.} Accordingly, the human face based on NeRFBlendShape is rendered by
\begin{align}
    \hat{\boldsymbol{I}}=V(\mathcal{R}_\beta\left(C, \boldsymbol{e}\right)),
\end{align}
where $V(\cdot)$ represents the volumetric rendering in \eqref{volumerendering}. In NeRFBlendShape, expression features in different levels of the hash table could be trained together, which could achieve swift face rendering in tens of milliseconds by avoiding frequent querying.
\vspace{-3pt}
\subsection{Exploiting face NeRF model as Semanticl Knowledge Base}
Motivated by the success of the face model and NeRFBlendShape, we propose to utilize the NeRFBlendShape-based face NeRF model as the semantic knowledge base to facilitate semantic communications of 3D human faces. In particular, semantic knowledge base aims to provide rich knowledge for semantic information processing (i.e., feature extraction and data recovery), which is essential in semantic communication systems \cite{ren2023knowledge}. Generally, the semantic knowledge base could be pre-trained beforehand in an off-line manner \cite{lyu2024rethinking}, and then be shared at the transmitter and receiver without influencing the communication latency in real time.

The employment of NeRFBlendShape as the facial knowledge base has the following advantages. On the one hand, it implicitly stores the complete personalized 3D face information, which can help to reconstruct the raw face with semantic features. On the other hand, NeRFBlendShape helps resolve the enormous time consumption for training and rendering of NeRF. 

In this work, we pre-train the NeRFBlendShape-based facial knowledge base with a segment of monocular video of human faces as the training dataset in an offline manner. By delivering the pre-trained knowledge base to the receiver, the transmitter and receiver can share the common face NeRF model. Notice that the semantic knowledge base corresponds to the facial identity, and as a result, the knowledge base only needs to be transmitted once for the same person over a long time period consisting of a large number of video chunks. 

\subsection{NeRF-SeCom Workflow} \label{workflow}

\begin{figure*}
    \centering    \includegraphics[width=0.57\linewidth]{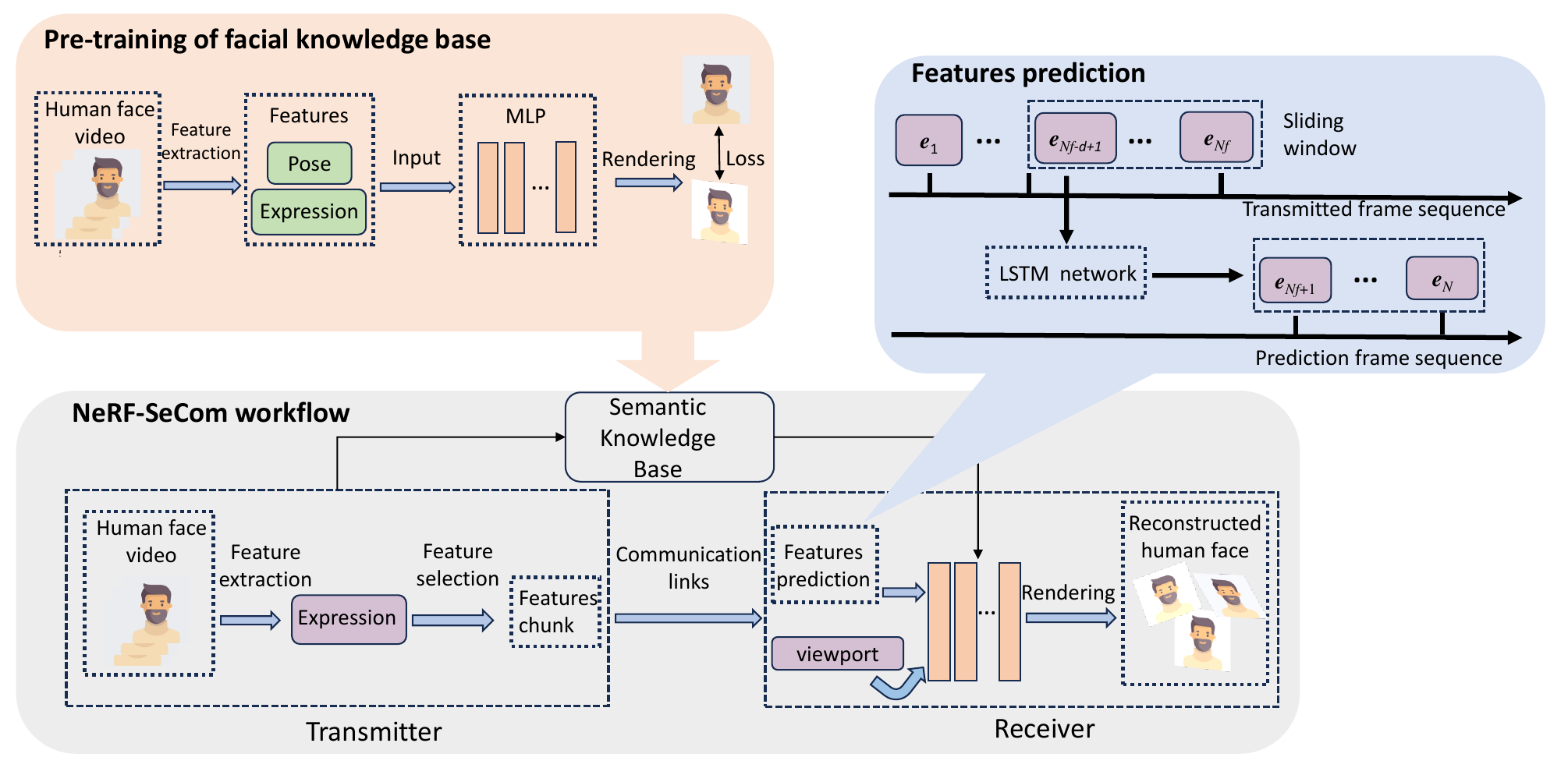}
    \caption{Illustration of NeRF-SeCom, the proposed semantic communications framework for 3D human face transmission with NeRF.}
    \label{SemanticFramework}
    \vspace{-5pt}
\end{figure*}

In this subsection, we present the workflow of our proposed NeRF-SeCom framework, as shown in Fig.~\ref{SemanticFramework}. 

\subsubsection{Semantic Feature Extraction}
First, semantic feature extraction is performed on the input video at the transmitter. The face feature extraction follows from the blendshape method \cite{Cao6654137FaceWarehouse}, in which each expression coefficient has a corresponding specified semantic meaning, such as eye closing and mouth opening. Specifically, this scheme takes the video frames as input and outputs the expression features and pose features of the human face frame by frame. Therefore, we obtain the expression coefficients and the face pose parameters of $n$-th frame as
\begin{align}
 \left(\boldsymbol{e}_{n}, p_{n}\right)=\mathcal{T}({\boldsymbol{I}}_n),
\end{align}
where $\mathcal{T}(\cdot)$ denotes the features extraction scheme and $p_{n}$ denotes the corresponding pose parameters. Similar to \cite{Hong_2022_CVPR,Gao3555501nerfface}, we take the human head pose parameters as camera parameter $C_n$. Then, $C_n$ is used for the ray marching to obtain camera rays and sampling points along those rays during the training process. For inference, viewers can render according to their viewing directions. In such a case, we focus on the selection and transmission of expression feature parameters in subsequent modules, and the viewing directions (i.e., viewport) of users are known at the receiver.

\subsubsection{Feature Chunk Packing} Next, features are packed into a video chunk. Unlike traditional video streaming, which encodes a batch of video frames into video chunks for each transmission, our framework packs the extracted  features into chunks for transmission. To efficiently reduce the communication overhead for transmission, we classify the expression features into static and dynamic types. Accordingly, for each chunk, we allow the transmitter to send the average value of static expression coefficients only once for all the frames, and transmit the dynamic expression coefficients for a number of $N_f \leq N$ frames to adhere to the rate constraint. In particular, suppose that the total number of expression features is $M$ and that of dynamic expression features is $M_{\text{dyn}}$. Consider that each expression coefficient in one frame is quantized to a size of $Q$ bits. Therefore, the transmitted bits in the first frame are $Q M$, and the transmitted bits in the remaining frames are $Q(N_f-1) M_{\text{dyn}}$. Then we have the total bits to be transmitted as $QM + Q(N_f-1)M_{\text{dyn}}$, which should not exceed $\tau R$, with $\tau$, $R$ denoting the latency and communication rate, respectively. Accordingly, we have
\begin{align} 
    Q M+ Q(N_f-1) M_{\text{dyn}} \leq \tau R,
\end{align}
Accordingly, it follows that the number of frames with dynamic feature transmission $N_f$ is given by 
\begin{align}
    {N_f}=\frac{{\tau R}}{{Q M_{\text{dyn}} }}{\rm{ - }}\frac{M}{M_{\text{dyn}}}{\rm{ + }}1. \label{calculateframe}
\end{align}
Notice that in the feature chunk packing, it is essential to select the expression features, which will be discussed in detail in Section \ref{selecttion}. 

\subsubsection{Face Rendering} After receiving the  expression features in the whole chunk, the receiver renders the 3D human face by utilizing the NeRF models in the shared facial knowledge base. Here, the users can freely change their viewing directions to render the viewpoints of the 3D content. However, the face rendering needs the complete expression features over all frames, but only dynamic expressions in the first $N_f$ frames are transmitted. Therefore, we need to develop proper feature prediction methods for estimating the dynamic expressions in the remaining $N - N_f$ frames based on the received features. In particular, we develop a long short-term memory (LSTM) network for prediction, as will be discussed in Section \ref{predictionLSTM}.


\section{Semantic Feature Selection and Prediction} \label{ProposedSchemes}
This section presents our proposed semantic feature selection and prediction mechanisms, which enhance efficiency of feature chunk packing by determining dynamic features, and ensures predict features that are not transmitted, respectively.
\subsection{Feature Selection at Transmitter} \label{selecttion}
In our proposed NeRF-SeCom, facial expressions are depicted by high-dimensional expression coefficients. In practice,  many expression coefficients in actual facial videos remain unchanged or change slightly over certain periods. In this case, the contribution of slightly changing coefficients to the morphological changes of the entire face is small, showing low importance in a video sequence. Therefore, transmitting these static (or nearly static) expression coefficients in each independent frame incurs a waste of communication resources. To reduce communication overhead while ensuring the quality of facial reconstruction, we propose an expression selection scheme to decompose the static expression and the dynamic expression features by taking into the variance of the values of features in a duration.

In particular, we consider a duration as $N$ frames for video transmission. The $m$-th extracted expression feature coefficients during $N$ frames are $\boldsymbol{E}_m=\left\{e_m^1,e_m^2,\ldots,e_m^N\right\}$, $\forall m \in \left\{1,\ldots,M \right\}$. Accordingly, the average value and the variance of this expression feature during $N$ frames are given by
\vspace{-3pt}
\begin{align}
        \bar{e}_m^N=\frac{1}{N} \sum_{t=1}^{N} e^t_{m},~
        D^2(\boldsymbol{E}_m)=\frac{1}{N} \sum_{t=1}^{N}\left(e_m^t-\bar{e}_m^N\right)^2. 
\end{align}
With the variance values $D^2(\boldsymbol{E}_m)$ in hand, We categorize the expression features into dynamic and static features as follows. In specific, if the variance exceeds a given threshold $\delta$, it indicates that the expression coefficient varies dynamically over the $N$ frames and significantly contributes to the change in the entire face; otherwise, it is considered static. We denote the set of dynamic expression coefficients as $\mathcal{D}$ and the set of static expression coefficients as $\mathcal{S}$, respectively. In this case, we use an indicator function to define whether the $m$-th ($\forall m \in \left\{1,\ldots,M \right\}$) expression feature is dynamic, which is given by
\begin{align}
    {\varPsi(\boldsymbol{E}_m)} = \left\{ \begin{array}{l}
      1,~~\text{if}~D^2(\boldsymbol{E}_m) \geq \delta,\\
      0,~~~~\text{otherwise}.
      \end{array} \right.
\end{align}
Accordingly, we calculate the number of dynamic expression features $M_{\text{dyn}}$ by $M_{\text{dyn}}= \sum_{m=1}^M \varPsi(\boldsymbol{E}_m)$. Based on $M_{\text{dyn}}$, we can determine the frames that can be transmitted during the feature chunk packaging phase (see \eqref{calculateframe}). 

\subsection{Feature Prediction at Receiver} \label{predictionLSTM}
To meet the communication rate requirement, the dynamic features are only transmitted over a small portion of $N_f$ frames over the whole of $N$ frames with $N_f<N$. In such a case, the receiver can only render parts of the whole 3D human face based on the received features, leading to the deterioration of the face reconstruction in 3D immersive video.

Padding the average value or the last value of expression coefficients to the non-transmitted frames is an intuitive method to recover all frames in the chunk and provide a stable frame rate for video streaming. However, this method does not take into account the changes in features, making it difficult to ensure the quality of experience. To ensure the visualization quality at the receiver, we develop an LSTM-based feature prediction scheme. For the dynamic expression bases, we predict the non-transmitted parts in the remaining $N-N_f$ frames of each chunk. For the static expression features, we pad the average value to the remaining frames. Specifically, we consider a sliding window with step size $d$ to predict the expression coefficients of the face over time, as shown in Fig. \ref{SemanticFramework}. The dynamic expression coefficients in $i$-th frame $e_m^i$ ($\forall i \in \left\{N_f+1,\ldots,N\right\},~\forall m\in \mathcal{D}$) are predicted based on the past $d$ status $e_m^{i-d:i-1}=\left\{e_m^{i-d},e_m^{i-d+1},\ldots,e_m^{i-1}\right\}$, which can be written as
\begin{align}
    e_m^i \!= \!\mathcal{L}_{\alpha_m}(e_m^{i-d:i-1}),~\forall i \in \left\{N_f+1,\ldots,N\right\},
\end{align}
where $\mathcal{L}_{\alpha_m}$ is the LSTM prediction networks with learnable parameter $\alpha_m$ obtained from the training of $m$-th expression coefficient. For the static expression features, we have 
\begin{align}
    e_m^i = e_m^1,~\forall i \in \left\{1,\ldots,N\right\},~\forall m\in \mathcal{S}.
\end{align}
Once the prediction is completed, we possess the full feature information in one chunk for 3D face rendering in Section \ref{workflow}.

\section{Experimental Results} \label{experiment}

This section presents numerical results to validate the performance of the proposed NeRF-SeCom framework. Specifically, in the experiments, we set the total dimension of expression coefficients as $M=47$, the fixed maximum amount of frame in one chunk as $N=100$, the quantization level of each expression coefficient as $Q=16 \textup{bits}$, the latency constraint as $\tau=1\textup{ms}$, the variance threshold as $\delta=0.01$, and the step size of LSTM network as $d=5$, respectively.  

We consider the following benchmark schemes for performance comparison:
(1) \textbf{NeRFBlendShape}: The features of all $N$ frames are accurately transmitted without any rate limitations. This serves as the performance upper bound. (2) \textbf{NeRF-SeCom without feature selection}: Instead of using feature selection module, both static and dynamic expression coefficients are transmitted. Accordingly, the total number of frames with feature transmission is $\hat{N}_f={\tau R}/{QM}$. The expression coefficients of the remaining $N- \hat{N_f}$ frames are predicted as $\hat{e}_m^i= \mathcal{L}_{\alpha}(e_m^{i-d:i-1})$, with $\forall i \in \left\{N_f+1,\ldots,N\right\},~\forall m\in \left\{1,\ldots,M\right\}$.
(3) \textbf{NeRF-SeCom without feature selection and prediction}: We deactivate the feature selection and prediction modules. The total number of frames to be transmitted is $\tilde{N}_f={\tau R}/{QM}$. The expression coefficients of the remaining $N- \tilde{N_f}$ frames are directly padded with that of the last frame, i.e., $\tilde{e}_m^i= e_m^{N_f}$, with $\forall i \in \left\{N_f+1,\ldots,N\right\},~\forall m\in \left\{1,\ldots,M\right\}$.
\begin{figure}
    \centering
    \includegraphics[width=0.7\linewidth]{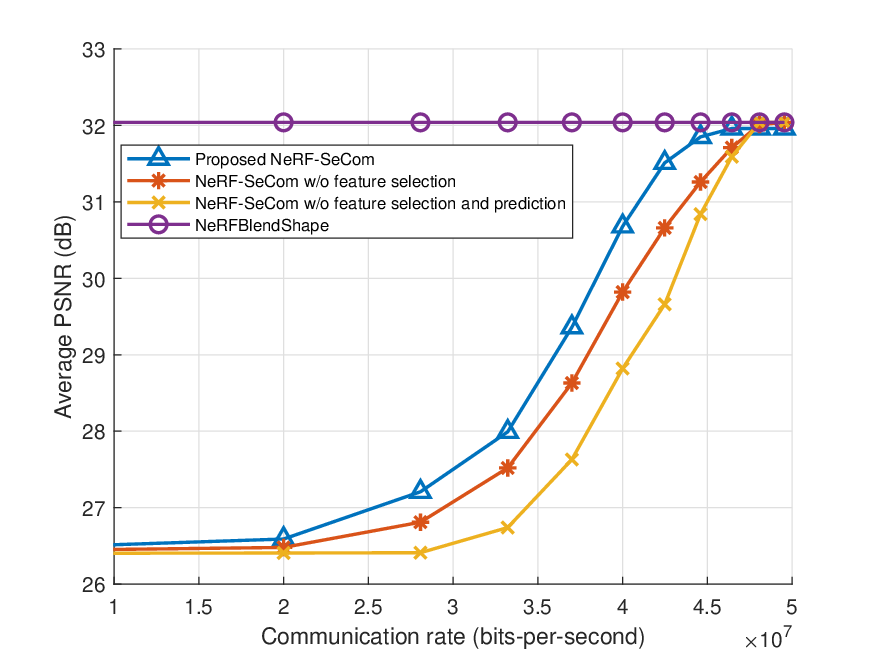}
    \vspace{-10pt}
    \caption{Performance comparison of our proposed NeRF-SeCom framework and benchmark schemes in terms of average PSNR under variational communication rate.}
    \label{PSNR}
    \vspace{-10pt}
\end{figure}
\begin{figure}
    \centering
    \includegraphics[width=0.95\linewidth]{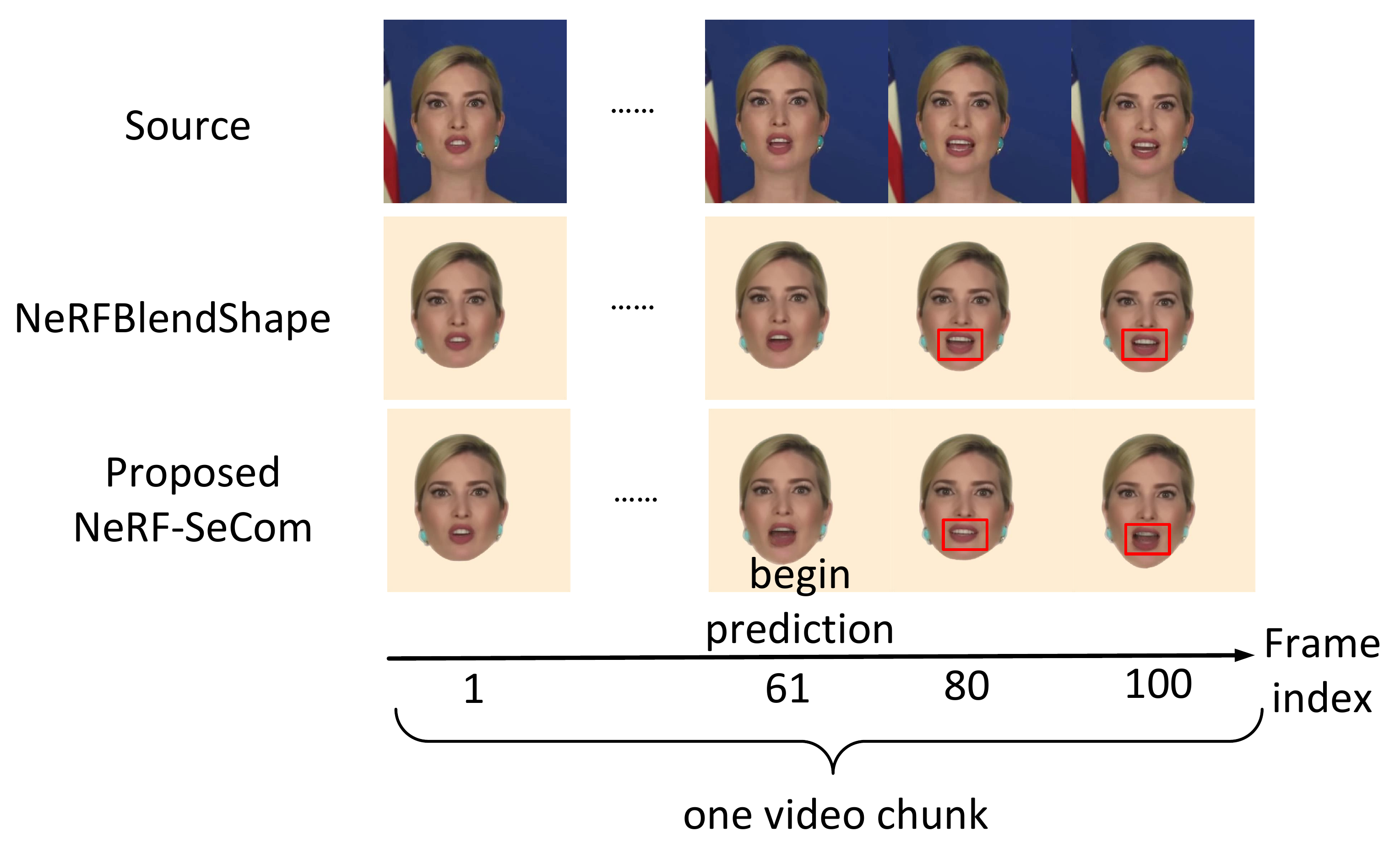}

    \caption{Rendering results comparison sampled from frame $61$, $80$, and $100$ with NeRFBlendShape and ground truth by setting $N_f=60$.}
    \label{semantic}

\end{figure}

Fig.~\ref{PSNR} shows the performance of our proposed NeRF-SeCom framework and the benchmark schemes under variational communication rates. In this case, we evaluate the performance via the average peak signal-to-noise ratio (PSNR) in one chunk (i.e., 100 frames). For each frame, PSNR is defined as
\begin{align}
    {\rm PSNR}(\boldsymbol{I},\boldsymbol{\hat I})= 10 {\rm log}_{10}( \frac{{\rm MAX}^2}{\frac{1}{3HW}\|\boldsymbol{I}-\hat{\boldsymbol{I}}\|^2}),
\end{align}
where ${\rm MAX}$ denotes the maximum possible value of the image
pixels. First, it is observed from Fig.~\ref{PSNR} that our proposed framework maintains satisfied reconstruction performance under different data rates, and the performance increases as the rate increases. Moreover, in high rate regime, the performance of our proposed framework converges to the upper bound since the rate can ensure the full transmission of total video frames in one chunk. Next, it is also observed that the proposed framework outperforms the benchmark schemes significantly, which demonstrates the effectiveness of our proposed framework with feature selection and prediction. However, in high and low rate regimes, the performance gaps between the proposed framework and benchmark schemes are limited. This is due to the fact that, on the one hand, in low rate regimes, long time-step predictions lead to accumulated errors. On the other hand, in high rate regimes, the need for feature selection and prediction significantly reduces due to the sufficiently high rate for feature transmission.

Fig.~\ref{semantic} shows the comparison of visual rendering results between our proposed framework, NeRFBlendShape, and the source image under $N_f=60$. We provide the rendering results of the 61-st frame (the first frame to start prediction), the 80-th, and the 100-th frame (the last frame) for comparison. It is observed that as the amount of predicted frames increases, the error of the rendered face rises. This is due to the fact that each shift of the sliding window incorporates the current predicted results as true values for subsequent predictions. In such a case,  the accumulation of prediction errors from the previous sequence compromises the system performance.

\section{Conclusion} \label{conclusion}

This paper proposed a semantic communication framework for 3D human face reconstruction with NeRF (NeRF-SeCom), which realized face reconstruction over communication networks based on NeRF human face models. We proposed to transmit the features of faces rather than the raw 3D face information, and developed feature selection and prediction modules to resist the effect of variational communication rate. Numerical results were provided to demonstrate the effectiveness of our proposed framework. We hope that this paper can pave the way to the efficient 3D content transmission via NeRF. In the journal version, we have further extended our work to the training and inference of NeRF over 6G networks \cite{wu2024embracing}.

\bibliography{IEEEabrv,reference}

\bibliographystyle{IEEEtran}

\end{document}